\documentclass[prd,showpacs,preprintnumbers,floatfix,superscriptaddress,10pt]{revtex4}
\usepackage{color}
\newcommand{\de}{\partial}

\newcommand{\be}{\begin{equation}}
\newcommand{\ee}{\end{equation}}
\newcommand{\ba}{\begin{eqnarray}}
\newcommand{\ea}{\end{eqnarray}}

\begin{document}
\title{Transport theory for  cold  relativistic superfluids from an analogue model of  gravity}
\author{Massimo Mannarelli and Cristina Manuel}
\address{$^1$ Instituto de Ciencias del Espacio (IEEC/CSIC) \\
Campus Universitat Aut\` onoma de Barcelona,
Facultat de Ci\` encies, Torre C5 \\
E-08193 Bellaterra (Barcelona), Spain}
\date{\today}

\begin{abstract}
We write a covariant transport equation for the phonon excitations  of a relativistic superfluid  valid at small temperatures. The hydrodynamical equations for this system are derived from the
effective field theory associated to the superfluid phonons. We describe how to 
construct the kinetic theory for the phonon quasiparticles using a relativistic generalization
of the analogue model of gravity developed by Unruh.
This gravity analogy relies on the equivalence between the action of a phonon field moving in a superfluid background with  that of
a boson propagating in a given curved space-time. Exploiting  this analogy we obtain
continuity equations  for the phonon current, entropy and  energy-momentum tensor in a covariant form,   valid in any reference frame. Our aim is to shed light on some aspects of transport phenomena of relativistic superfluidity.  In particular, we are interested  in  the low temperature regime of the color flavor locked phase, which is a color superconducting  and superfluid phase of high density QCD that may be realized in the core of neutron stars.
\end{abstract}
\maketitle

\section{Introduction}
Superfluidity is a phenomenon that occurs in some systems at sufficiently low temperatures  after the appearance of
a quantum condensate \cite{landaufluids,IntroSupe}. 
Landau gave an explanation of this effect and developed his famous two-fluid model considering that the system
can be viewed as composed by  a superfluid component, which describes the coherent motion of the condensate, and a normal fluid component, where dissipative processes are allowed. 
The property of superfluidity follows from the existence of some elementary excitations with a linear dispersion law,
the phonons. 
These phonons  are the Goldstone modes which originate from the spontaneous breaking of a global symmetry caused  
by the appearance of the condensate.

In the cold regime dissipative processes  are mainly due to the collisions of phonons, 
which are part of the normal fluid component. The viscous hydrodynamics of a superfluid is involved because one can define 
more transport coefficients than in a normal fluid. However, for non-relativistic superfluids one can compute from the kinetic theory constructed by Khalatnikov \cite{IntroSupe} all the transport coefficients that enter into the two-fluid hydrodynamical equations.


Relativistic generalizations of Landau's two-fluid model of superfluidity have been developed
for applications to neutron stars physics. These generalizations are rather non-trivial,
and  alternative formulations of the hydrodynamics have been proposed in the literature 
 \cite{khalatnikov,Carter,Carter-Kha,Carter-Lang,Son:2000ht,Son:unpubl}.
In the non-dissipative limit it is possible to show that
all these  approaches are equivalent.
Unfortunately, very little is known about dissipative effects in relativistic superfluids
\cite{Pujol:2002na, Gusakov:2007px,Valle:2007xx}. To the best of our knowledge, the whole set of
dissipative terms which enter into the relativistic hydrodynamical equations has only been derived in one of the above mentioned approaches \cite{Pujol:2002na, Gusakov:2007px}.

It is our aim to formulate the kinetic theory associated to the phonons of
a relativistic superfluid, as these quasiparticles should dominate the transport phenomena
at low temperatures. While the dynamics of the phonons in the frame where the superfluid
component is at rest is easily formulated, this is in principle not so when done in an arbitrary frame. Relativistic boosts of the different variables complicate
in an unnecessary way the problem. However, there is still the possibility to obtain transport equations 
in a fully covariant way. The approach 
exploits the gravity analogs which are naturally associated
to relativistic hydrodynamical waves \cite{Bilic:1999sq}, following ideas of Unruh \cite{Unruh} and others (see Ref.\cite{Barcelo:2005fc} for a review and references).
 
Let us recall here that Unruh established long ago an analogy between supersonic flow in hydrodynamics
and black hole physics
\cite{Unruh}.
Exploiting this hydrodynamics/gravity analogy \cite{Stone:1999gi},
Volovik pointed out that the kinetic equation of the phonons of a non-relativistic superfluid
can be rewritten as that of a boson evolving in Unruh's acoustic metric \cite{Volovik:2000ua}.
It is thus natural to expect that the phonons of a relativistic superfluid admit a similar
treatment, as we show here. Let us mention that Popov also derived a similar transport equation
for the superfluid phonons \cite{Popov:2006nc}. We will implement the same ideas in a somewhat different way.

Relativistic superfluids may be realized in nature in the interior of neutron stars where the temperature is low and the energy scale of the particles is sufficiently high.   In particular, in the inner crust of neutron stars the attractive interaction between neutrons can lead to the formation of a BCS condensate, the system then becoming superfluid.
Moreover, if deconfined quark matter is present in the core of neutron stars
it will very likely be in a color superconducting phase \cite{reviews}.
Quantum Chromodynamics (QCD) predicts that
at asymptotically high densities  quark matter is in the color flavor locked  phase 
(CFL)~\cite{Alford:1998mk}.
In this phase up, down and strange quarks pair forming a quark condensate that
is antisymmetric in color and flavor indices.
 The  order parameter breaks the baryonic number $U(1)_B$
symmetry spontaneously, and therefore CFL quark matter is  a superfluid as well.

The main part of the present paper is dedicated to  the CFL superfluid. The starting point of our analysis is the low energy effective theory of the CFL phase which is obtained from the equation of state of quark matter \cite{Son:2002zn}. 
Then we show how   at asymptotic high density it is possible to derive from the
microscopic physics, from QCD, the hydrodynamical behavior of the  CFL superfluid. 
 
Our ultimate goal is having a well-defined formalism to derive all the transport
coefficients associated to the CFL superfluid.
 Up to now only the shear viscosity and bulk viscosity associated to the phonon contribution
of the CFL superfluid have been computed \cite{Manuel:2004iv,Manuel:2007pz}, as well as the
contribution to the bulk viscosity due to kaons \cite{Alford:2007rw}, and an estimate of the
thermal conductivity \cite{Shovkovy:2002kv}.
However there are other transport coefficients which are peculiar of superfluid systems that
are still unknown. These are needed in order to derive the macroscopic behavior of a
hypothetical compact star made of  CFL quark matter, and single out possible signatures of
quark matter in astrophysical scenarios.

This paper is organized as follows.
In Section \ref{Sec-hydro} we review the convective variational approach of
relativistic superfluid hydrodynamics.  In Section \ref{Sec-CFL} we review the derivation of the hydrodynamical equations for the CFL superfluid phase of QCD and the low energy effective action for the superfluid  phonons.
In Section \ref{Sec-kinetic} we present the kinetic theory for the phonons of a relativistic superfluid and derive the continuity equations for the phonon current, entropy and energy-momentum tensor.  In Section \ref{Sec-thermal} the transport equations are specialized to a local equilibrium state and the corresponding  expressions of number density, entropy density and pressure are computed.
We show that in the limit of small velocities these quantities agree with those of the phonons of
a non-relativistic superfluid.
 We draw our conclusions in Section \ref{Discussion}. 
In Appendix \ref{section-PB}, we review an hydrodynamical approach to relativistic superfluidity 
by Son.
We will  assume a  physical Minkowski space-time metric (there is no real gravity field)
and use the conventions  $\eta^{\mu\nu} =(1,-1,-1,-1)$, with natural units  $c= \hbar =k_B=1$.

\section{Hydrodynamics of relativistic superfluids: convective variational approach}
\label{Sec-hydro}
There are different formulations of the hydrodynamical equations governing  a relativistic superfluid
\cite{khalatnikov,Carter,Carter-Kha,Carter-Lang,Son:2000ht,Son:unpubl}. All of them were derived as
relativistic generalizations of the equations of  Landau's two-fluid model of non-relativistic superfluid dynamics \cite{landaufluids,IntroSupe}.
Here we review the convective variational approach  of Carter \cite{Carter}. It has been shown in Ref.\cite{Carter-Kha} that this approach is equivalent, at least in the non-dissipative limit, to the potential variational approach  developed by Lebedev and Khalatnikov \cite{khalatnikov}.
We also present in  Appendix \ref{section-PB} the formulation derived by Son, which is  equivalent to
Carter's formulation as well.

In the convective variational approach one defines a master thermodynamic function $\Lambda$,  which is a scalar function of  three independent scalar variables, namely $n^\rho n_\rho, s^\rho s_\rho$ and
$n^\rho s_\rho$, where $s^\rho$ is the entropy current vector  and $n^\rho$ is the  (total) particle number vector.
All the thermodynamic equations  of the superfluid can be derived from 
$\Lambda$ and by partial differentiation one finds 
\begin{equation}
d \Lambda = \Theta_\rho ds^\rho + \mu_\rho d n^\rho \, ,
\end{equation}
that serves to define the so-called thermal momentum covector  $\Theta_\rho$ and the vector  $ \mu_\rho$.

By applying a Lengendre-type transformation, one can define the thermodynamics in terms
of the pressure
\begin{equation}
P = \Lambda - s^\rho \Theta_\rho - n^\rho \mu_\rho \,,
\end{equation}
which  is a function  of the conjugate variables $\mu^2 = \mu_\rho \mu^\rho$, $z^2 = \mu_\rho \Theta^\rho$
and $\Theta^2 = \Theta^\rho \Theta_\rho$. Here $\mu$ is the chemical potential of the system, and
$\Theta$ represents the temperature, which we will also denote by $T$.
  By partial differentiation one has
\begin{equation}
dP = s^\rho d\Theta_\rho + n^\rho d\mu_\rho \ ,
\end{equation}
and then one can express the vectors    $n^\rho$ and $s^\rho$ as a function of $\mu^\rho$  and $\Theta^\rho$ as
\begin{equation}
n^\rho = F \mu^\rho + Q \Theta^\rho    \ , \qquad s^\rho = Q \mu^\rho + G \Theta^\rho \ ,
\end{equation}
where
\begin{equation}
F = 2 \frac{\partial P}{\partial \mu^2} \ , \qquad Q =  \frac{\partial P}{\partial z^2} \ , \qquad
G = 2 \frac{\partial P} {\partial \Theta^2} \ .
\end{equation}

Finally  the energy-momentum tensor of the system is given  by
\begin{equation}
\label{Sup-enermom}
T^{\rho \sigma} = n^\rho \mu^\sigma + s^\rho \Theta^\sigma - P \eta^{\rho \sigma} \,.
\end{equation}
Although the notation used does not demonstrate the symmetry
of the tensor explicitly, it really takes place \cite{Carter-Kha}.

The hydrodynamical equations of  the superfluid are given by 
\begin{eqnarray}
\label{sup-hydroeqs}
&\partial_\rho n^\rho   =  0  \ , & \qquad
\partial_\rho s^\rho   =   0 \ , \\
& s^\rho \partial_{[\rho} \Theta_{\sigma ]}  =  0  \ , & \qquad \partial_{[\rho} \mu_{\sigma]}  = 0 \ ,
\end{eqnarray}
where the brackets denote index antisymmetrization. The irrotational condition equation on $\mu_\rho$ tells that this vector can be expressed as the gradient of a scalar function,
\begin{equation}
\label{sup-covect}
\mu_\rho = -\partial_\rho \varphi \ .
\end{equation}
As it will be shown in the following Section, for the CFL superfluid
it is possible to show that this scalar function is related to the phase of the  diquark condensate
\cite{Son:2002zn}.

Using  Eq.~(\ref{Sup-enermom}) and  Eqs.~(\ref{sup-hydroeqs}) one obtains the energy-momentum conservation law
\begin{equation}
\partial_\rho T^{\rho \sigma} = 0 \ .
\end{equation}
From the expressions above it is clear that  the hydrodynamical equations keep the form of conservation laws, exactly as in
a normal fluid.

In the zero temperature limit and for the ideal case of no interaction between the particles,  the pressure is only a function of the chemical potential, and  the entropy current vanishes.
In this case, one defines the velocity vector
\begin{equation}
v^\rho = \frac{\mu^\rho}{\mu} \ ,
\end{equation}
such that it is properly normalized, $v^\rho v_\rho = 1$.
In the $T=0$ limit
the energy-momentum tensor takes the form of that of an ideal fluid and  then Eq.~(\ref{Sup-enermom})
 can be written as
\begin{equation}
T^{\alpha \beta} = ( n \mu) v^\alpha  v^\beta - P \eta^{\alpha \beta} =  \left(\rho+P\right) v^\alpha  v^\beta -
 P \eta^{\alpha \beta} \,,
\end{equation}
where $\rho$ is the energy density of the system, and we have used the zero temperature relation
 $n \mu = \rho+P$.

Unfortunately in this formulation of the superfluid hydrodynamics it is not a priory obvious 
what is the physical meaning of all the hydrodynamical variables. 
The approach presented in this paper allows one to clarify 
their field-theoretical origin.

\section{The CFL Superfluid in the  limit of vanishing temperature}
\label{Sec-CFL}

In this Section we  study the  relativistic superfluid system composed of  color-flavor locked (CFL) quark matter. The CFL   phase has long been known to be a superfluid:  by picking a phase
its order parameter breaks the  baryonic number $U(1)_B$
symmetry spontaneously. At least in the
limit of asymptotic high densities this is the energetically favored phase with respect to other less symmetric  color superconducting phases and with respect to unpaired quark matter \cite{reviews}.  

In the regime of  sufficiently high densities it is possible to derive
the hydrodynamical equations from the microscopic physics.
Indeed in  Ref.~\cite{Son:2002zn} Son showed how to obtain the effective action of the superfluid phonons  by integrating out from the QCD Lagrangian the heavy degrees of freedom corresponding to gluons, quarks and mesons.
At very high densities, this is an allowed  operation because the only low energy excitations are the phonons. 
The classical equations of motion derived from this effective action
correspond to the zero temperature hydrodynamical equations that can be obtained within the
convective approach, discussed in Section~\ref{Sec-hydro}. This way of obtaining the Carter's equations allows one to  identify
the  relation between the superfluid velocity and the gradient of the phase of the diquark condensate. Indeed, these two quantities turn out to be directly proportional. 

In order to see thermal effects in the low temperature limit in full generality,
one has to  consider the dynamics of the phonons moving in the superfluid background. This can  also be
derived from  Son's approach \cite{Manuel:2007pz}, and 
we discuss in Section~\ref{Phonons-moving}  how to obtain the
zero temperature effective action of the phonons in the presence of the superfluid background. As announced  in the Introduction, the equation of motion of the phonon field in the  superfluid background can be interpreted as that of a scalar field moving in an ``acoustic" non-flat metric.
This means that   we will write the equation of motion in a covariant form with respect to such a non-flat metric. This formulation is rather useful, as then one can infer the phonon effective action in
any frame, and not necessarily in the superfluid rest frame. We will then use it in order to
obtain the transport theory in covariant form.

Let us stress here that we only consider the very low temperature regime, and that by increasing the temperature other quasiparticles can be excited in the CFL phase. Although such excitations  might be relevant  for transport phenomena at moderately higher temperature \cite{Alford:2007rw}, for the sake of 
simplicity we will ignore them, as we assume  a very low temperature.

\subsection{The effective Lagrangian of the superfluid phonons}
\label{Sec-CFL-Son}

The effective field theory  for the  Goldstone boson originating from the spontaneous breaking of the $U(1)_B$ symmetry can be constructed from the equation of state (EoS) of normal quark matter, following the procedure of
Ref.~\cite{Son:2002zn}. Because the superfluid phonon is a
Goldstone boson, and thus a light degree of freedom, its low
energy effective field theory is obtained by integrating out the
heavy modes from the QCD Lagrangian. This operation amounts to a
minimization procedure with respect to the modulus of the order
parameter of the symmetry. Exploiting other symmetries of the
problem, one can entirely fix the form of the phonon effective
action from the knowledge of the pressure of CFL quark matter.

Calling  $\varphi$  the phase of the
condensate, and  defining
$D_\rho  \varphi \equiv \partial_\rho \varphi - \mu_q A_\rho$, with $A_\rho = (1,0,0,0)$ and $\mu_q$  the quark chemical potential,
 the low energy effective Lagrangian for $\varphi$ (valid for energy scales much smaller than the the CFL gap $\Delta$) is expressed as
\begin{equation}
{\cal L}_{\rm eff}[D_\rho \varphi] = P [(D_\rho \varphi D^\rho  \varphi)^{1/2}] \ .
\end{equation}
In this expression  $P$ is a functional of the derivatives of the field $\varphi$ that has the same form as the pressure of the system at zero temperature. Therefore in order to obtain the Lagrangian  for $\varphi$ is enough to specify $P$.
At asymptotic large densities  the EoS of CFL quark matter  takes the form 
\begin{equation}
 \label{freeEOS}
P (\mu_q)
=\frac{3}{4\pi^2} \mu_q^4 \, ,
\end{equation}
corresponding to a system of three flavors of  massless  quarks. The validity of such an expression relies on the fact that   at very high $\mu_q$  the coupling constant is small $g (\mu_q) \ll 1$ and  the effects
of interactions and the effects of Cooper pairing are subleading. Therefore they are  neglected in Eq.~(\ref{freeEOS}). One  assumes that  the quark masses give a subleading effect as well, because  $m_q \ll \mu_q$. Effects  due to the strange quark mass are considered in Ref.~\cite{Manuel:2007pz}.
In the end one obtains from Eq. (\ref{freeEOS}) that the low energy effective Lagrangian for the CFL phase in the limit of asymptotic densities is given by
\begin{equation}
\label{L-BGB-0}
{\cal L}_{\rm eff}  = \frac{3}{4 \pi^2}
\left[ (\partial_0 \varphi - \mu_q)^2 - (\partial_i \varphi)^2
\right]^2 \, .
 \end{equation}

There is an interesting  interpretation of the equations of motion
associated to $\varphi$. Since the Lagrangian in Eq.~(\ref{L-BGB-0}) does not explicitly depend on the field $\varphi$, but only on its derivatives, the corresponding classical equation of motion  takes the form of  a conservation law. Then one can view it  as the  hydrodynamical
conservation law of a current representing baryon number flow,
\begin{equation}
\label{S-hy-1}
 \partial_\nu (n_0 v^\nu) = 0 \ ,
\end{equation}
 where \be n_0 =\frac{dP}{d \mu} \Big |_{\mu =\bar \mu} = \frac{3}{\pi^2} \bar \mu^3 \ee is interpreted
as the baryon density \cite{Son:2002zn}, where  $\bar \mu = (D_\rho  \bar \varphi D^\rho  \bar \varphi)^{1/2}$ and where
\begin{equation} \label{svelocity}
 v_\rho = - \frac{D_\rho  \bar\varphi}{\bar \mu} \, ,
\end{equation}
is the superfluid velocity  with $\bar \varphi$ the solution of the classical equation of motion. 
Notice that the superfluid velocity is properly normalized, that is,  $v_\rho v^\rho = 1$, and that in this definition it is taken into account that Lorentz symmetry is explicitly broken by $\mu_q$.

The energy-momentum tensor  can  be written in terms of the velocity
defined in Eq.~(\ref{svelocity}) and Noether's energy-density $\rho_0$,
 \begin{equation}
\label{S-hy-2}
T^{\rho \sigma}_0 = (n_0 \bar \mu)  v^\rho  v^\sigma - g^{\rho \sigma} P_0 = (\rho_0 + P_0)  v^\rho  v^\sigma - \eta^{\rho \sigma} P_0 \ ,
 \end{equation}
where we have written  $\rho_0 + P_0 = n_0 \bar \mu$, with $P_0$  the quark pressure evaluated at
$\bar \mu$.
 The  energy-momentum tensor  is conserved
\begin{equation} \label{Tconserved}
\partial_\rho T^{\rho \sigma}_0 = 0\ ,
\end{equation}
and traceless $ T^\rho_{0 \,\rho}=0$.

One can immediately check that Eqs.~(\ref{S-hy-1}) and (\ref{Tconserved}) are equivalent to the hydrodynamical equations of a relativistic superfluid
at zero temperature obtained in the convective variational approach. The definition of the
superfluid velocity differs in the zero component, but this can be matched by the change
$\varphi \rightarrow \varphi + \mu x_0$.

\subsection{Phonons moving in the superfluid background}
\label{Phonons-moving}

From the expression of the  Lagrangian in Eq.~(\ref{L-BGB-0}) it is possible to derive the effective field theory of the phonons moving in the background of the superfluid \cite{Manuel:2007pz}. 
In the CFL phase the superfluid phonon is the Goldstone boson associated to the breaking of the $U(1)_B$ symmetry and it
can be introduced  as the phase of the diquark condensate. However, according to Eq.(\ref{svelocity})
the gradient of the phase of the condensate  defines the superfluid velocity as well. Then it should  be possible to decompose the 
phase of the condensate in  two fields, the first describing the hydrodynamical variable, the second describing the quantum fluctuations associated
to the phonons. Therefore in order to find the phonon dispersion relation in a moving superfluid we will  consider the quantum fluctuations around the classical solution $\bar \varphi$ of the equations of motion associated to the Lagrangian (\ref{L-BGB-0}), thus we write
\begin{equation}
\label{split}
\varphi (x) = \bar \varphi(x) + \phi(x) \ .
\end{equation}
This splitting  implies a separation of scales -  the background field $\bar \varphi(x)$
is associated to the long-distance and long-time scales, while  the fluctuation $\phi(x)$ is
associated to rapid and small scale variations. The  gradient 
of $\bar \varphi(x)$ is proportional to the  hydrodynamical velocity, according to Eq.(\ref{svelocity}),
while the fluctuation $\phi$ is identified with the superfluid phonon.

From the low energy effective action of the system
\begin{equation}
S[\varphi] = \int d^4 x \, {\cal L}_{\rm eff}[\partial \varphi] \,,
\end{equation}
we deduce the effective action for  the phonon field  expanding around the stationary point corresponding to the classical solution
\begin{equation} \label{action2}
S[\varphi] = S[\bar \varphi] + \frac 12 \int d^4 x \,\frac{ \partial^2 {\cal L}_{\rm eff} }
{\partial(\partial_\mu\varphi) \partial(\partial_\nu \varphi)}
Á\Bigg \vert_{\bar \varphi}\partial_\mu \phi\, \partial_\nu \phi + \cdots \,,
\end{equation}
and considering  the expression of the   Lagrangian in Eq.~(\ref{L-BGB-0})  one has  that 
\begin{equation}
f^{\mu \nu} =  \frac{\partial^2{\cal L}_{\rm eff}}{\partial(\partial_\mu \varphi)
\partial(\partial_\nu \varphi)} \Bigg \vert_{\bar \varphi} = \frac{n_0}{\bar \mu}
\left \{ \eta^{\mu \nu} + \left(\frac {1}{c_s^2} - 1 \right) v^\mu v^\nu \right \} \ .
\end{equation}
Here $c_s = 1/\sqrt{3}$ is the
speed of sound in CFL quark matter.

The second term on the right hand side of Eq.(\ref{action2}) represents the  action of  the linearized fluctuation - here the superfluid phonon - 
and can be written as the action of a boson moving in a non-trivial gravity background \cite{Barcelo:2005fc}
\be
\label{phonon-action}
S[\phi] = \frac 12 \int d^4 x \sqrt{- {\cal G} } \, {\cal G}^{\mu \nu} \partial_\mu \phi\, \partial_\nu \phi \,,
\ee
where  we have defined the metric tensor
\begin{equation}
\label{phonon-metric}
{\cal G}^{\mu \nu} =   
\eta^{\mu\nu} + \left(\frac {1}{c_s^2} - 1 \right)  v^\mu  v^\nu 
\end{equation}
and   the determinant ${\cal G} = 1/{\rm det} | {\cal G}^{\mu \nu}|$. 

In writing the effective Lagrangian  in Eq.(\ref{phonon-action}) we have rescaled the phonon field according to
\be
\phi \to \sqrt{\frac{ c_s \bar\mu}{n_0}} \phi \,,
\ee
to get a dimensional scalar field, and thus a dimensionless metric in natural units. In doing the  rescaling, we have assumed that on the scale of variations of $\phi$, both $\bar \mu$ and $n_0$ can be assumed to be constant. This is justified by the separation of scales which is implicit in the splitting of the bosonic field $\varphi$ done in Eq.~(\ref{split}). 

The  normalized field obeys the  classical equation of motion
\begin{equation}
\partial_\mu \left( \sqrt{-  {\cal G}} \,  {\cal G}^{\mu \nu} \partial_\nu \phi \right) = 0 \,.
\end{equation}

We then see that from the effective  Lagrangian in Eq.~(\ref{L-BGB-0})
one can derive in Eq.~(\ref{phonon-metric}) the so-called sonic or acoustic metric tensor ${\cal G}^{\mu \nu}$ \cite{Carter-Lang,Barcelo:2005fc} that  depends on the speed of sound $c_s$ and on the superfluid velocity $v^\mu$. 
Note that
since the superfluid velocity $v^\mu$ is normalized to $1$, in general the metric tensor ${\cal G}^{\mu \nu}$  corresponds to a strong gravitational field. The limit of weak gravitational field is achieved for $c_s \to 1$, and only in this (unphysical) limit one can expand the gravitational metric tensor around the flat metric $\eta_{\mu\nu}$.

The dispersion law of phonons in the  superfluid medium are obtained by solving the  equation of motion for a massless particle moving in curved space time, that is 
\begin{equation}
\label{disp-eq}
{\cal G}^{\mu \nu} p_\mu p_\nu = 0 \ ,
\end{equation}
where $p_\mu = (E, -{\bf p})$.
In the superfluid rest frame, i.e. where $ v^\mu = (1,\bf 0)$, the dispersion law   simplifies to the form
\begin{equation}
E = c_s \, p \ ,
\end{equation}
which expresses the fact that phonons propagate at the speed of sound.
In a different frame the dispersion law takes a rather complicated form. In order to express such a relation in a compact way it is  convenient to define the following Lorentz invariants \cite{Popov:2006nc}
\begin{equation}
\varepsilon =  p_\mu v^\mu \ , \qquad \pi^2 = - \left(\eta^{\mu\nu} -  v^\mu  v^\nu \right) p_\mu p_\nu\,,
\end{equation}
which are the energy and the square of the 3-momenta in the local rest frame of the superfluid. It is easy
to check from Eq.(\ref{disp-eq}) that in any frame the relation $\varepsilon= c_s \pi$
holds.

\section{Kinetic theory for the superfluid phonons}
\label{Sec-kinetic}
In order to study the contribution of the superfluid phonons to the
hydrodynamics one has to consider a thermal bath of these quasiparticles, and
compute their contribution to the various thermodynamical quantities.
This analysis could be done in thermal field theory,
using the action given in Eq.~(\ref {phonon-action}).
Because we will be mainly interested in studying kinetic phenomena,
and ultimately in computing transport coefficients, we find more convenient
to develop a transport theory for the cold CFL superfluid. While it should be fully
equivalent to the thermal field theory approach, it turns out that the
computation of transport coefficients is much simpler when done with the
use of kinetic theory, as it is the case for other field theories (see, e.g. \cite{Litim:2001db}). 

Transport equations can only be formulated when there are
well-defined quasiparticles in the system, that is, thermal excitations that are sufficiently long lived.
In the CFL superfluid a direct computation of the damping rate for
the superfluid phonons shows that it is suppressed by $(T/\mu_q)^4$ with
respect to their typical energy \cite{Manuel:2004iv}. One can thus guarantee that it is possible to write a transport equation for these quasiparticles, as quark matter is in the CFL phase only when $T \ll \mu_q$.

\subsection{Quasiparticle picture}
As described in the previous Section,  the dispersion law of the phonon moving  
in the superfluid background can be seen as that of a boson propagating on
a gravity background characterized by the acoustic metric. In a classical approximation one
can view  the phonons as massless quasiparticles that
follow the paths determined by the geodesics of  the
acoustic metric, defined by  $0 = {\cal  G}_{\mu\nu} dx^\mu d x^\nu$. Moreover,  it is  possible to define a quasiparticle Hamiltonian, $ {\cal H}$, in such a way that solving the Hamiltonian
equations one obtains  the geodesics of the problem. Thus, we introduce
\begin{equation}
 {\cal H} =  \frac 1 2 {\cal G}^{\mu\nu}
 p_{\mu}  p_{\nu} = \frac 12 {\cal G}_{\mu\nu}   p^\mu  p^\nu = \frac 1 2 p^\mu  p_\mu \, ,
\end{equation}
where $p_\mu = {\cal G}_{\mu\nu} p^\mu$ and where   ${\cal  G}_{\mu\nu}$ is the inverse of the metric  ${\cal  G}^{\mu\nu}$, obtained by solving the equation
$ {\cal G}^{\mu\nu}{\cal  G}_{\nu\rho} = \delta^\mu_\rho$. For  the metric tensor defined in Eq.(\ref{phonon-metric}) the inverse is given by
\begin{equation} \label{inverse}
{\cal  G}_{\mu\nu}=  \eta_{\mu \nu} + \left(c_s^2 - 1 \right)  v_\mu  v_\nu \,,
\end{equation}
where $v_\mu = \eta_{\mu\nu} v^\mu$.

The Hamiltonian equations of motion are given by 
\begin{equation}
\label{H-eq}
\frac{d x^\mu}{d \tau} = \frac{\partial {\cal H}}{\partial  p_\mu} =
p^\mu\ , \qquad \frac{d  p_\mu}{d \tau} = - \frac{\partial {\cal H}}{\partial x^\mu}
= - \frac 12 \partial_\mu {\cal G}^{\alpha \beta}  p_\alpha  p_\beta \, ,
\end{equation}
however, since  we will formulate the transport approach using  the variables $(x^\rho, p^\rho)$,
rather than $(x^\rho,  p_\rho)$, it is convenient to find the equation of motion
for $p^\rho$. After simple algebraic manipulations one finds
\be
\label{tau-der-p}
\frac{d p^{\rho}}{d \tau} = -  {\cal G}^{\rho\mu}({\cal G}_{\mu\beta,\alpha}-\frac 1 2 {\cal G}_{\alpha\beta,\mu}) p^\alpha p^\beta = -\Gamma^\rho_{\alpha \beta} p^\alpha p^\beta \,,
\ee
where $\Gamma^\alpha_{\beta\gamma}$ denote the Christoffel symbols associated to
the metric ${\cal G}_{\mu \nu}$,
\begin{equation}
\Gamma^\rho_{\alpha\beta}= \frac 12  {\cal G}^{\rho \mu}
\left( {\cal G}_{\mu \beta,\alpha} + {\cal G}_{\mu \alpha, \beta}-  {\cal G}_{\alpha \beta,\mu} \right) \:.
\end{equation}
With  these Christoffel symbols one can define the  covariant derivatives for a generic controvariant vector $A^\mu$ as
\be
A^\mu_{;\nu} = A^\mu_{,\nu} + \Gamma^\mu_{\nu\alpha}A^\alpha \,,
\ee
where $A^\mu_{,\nu} = \partial_{\nu} A^\mu$.  While for a covariant vector one has
\be
 A_{\mu;\nu} =  A_{\mu,\nu}- \Gamma^\alpha_{\mu\nu}{ A}_\alpha \,.
\ee

\subsection{Covariant transport equation for the superfluid phonons}


We will now study the Liouville equation
governing the evolution of the  phonon distribution function $f(x,p)$. Here it is understood that  $f(x,p) p^\mu n_\mu  d \Sigma \,d {\cal P} $ is the number of particles whose world lines intersect the
hypersurface element $n_\mu d \Sigma$ around $x$, having four-momenta in the range
 $(p, p +dp)$,  with $n_\mu$  a four light-like vector that we will take as $v_\mu$,  and
the momentum measure $d {\cal P} $ is chosen in such a way that it is coordinate invariant \cite{Lindquist}
\begin{equation}
d {\cal P} = \sqrt{- \cal G} 2H(p) \delta({ \cal G}_{\mu \nu}p^\mu p^\nu) \frac{d^4p}{(2\pi \hbar)^3} \ ,
\end{equation}
where $H(p) = 1$ if $p$ is future oriented for an observer moving with velocity $v_\mu$, and $0$ otherwise.

The presence of the superfluid background will be manifest in the fact that
the Liouville equation has to be written in a covariant form with respect to the metric ${\cal G}^{\mu\nu} $. 
Indeed, the evolution of the distribution function is governed by the equation
\begin{equation}
\frac{d f}{d \tau} =  \frac{\partial
f}{\partial x^\alpha}\frac{d x^\alpha}{d \tau} +
\frac{\partial f}{\partial p^\alpha}\frac{d p^\alpha}{d \tau}
 = C[f] \, ,
\end{equation}
where $C[f]$ is the collision term and upon using  Eqs.~(\ref{H-eq},\ref{tau-der-p}) the Liouville equation can be rewritten as
\begin{equation}
\label{Blotzmman}
L[f] \equiv p^\alpha \frac{\partial f}{\partial x^\alpha} - \Gamma^\alpha_{\beta\gamma} p^\beta p^\gamma  \frac{\partial f}{\partial
p^\alpha}   = C[f]
\end{equation}
that is  the general relativistic version of the Boltzmann equation \cite{Lindquist}.

Knowing the distribution function one can construct the current, energy-momentum tensor and entropy associated to the phonons, which are given by the expressions

\begin{eqnarray}
\label{var-n}
n^\alpha_{\rm ph} & = & \int  p^\alpha f(x,p) d {\cal P} \ , \\
\label{var-T}
T^{\alpha \beta}_{\rm ph}  & = & \int p^\alpha p^\beta   f(x,p) d {\cal P} \ , \\
\label{var-S}
s^\alpha_{\rm ph}  & = &- \int   p^\alpha \left[ f \ln{f} - (1+f) \ln{(1+f)}\right] d {\cal P} \ .
\end{eqnarray}

Notice that in Section~\ref{Phonons-moving} we explicitly made a splitting of fields, separating the low energy modes from the high energy modes, the last corresponding to the phonons. However, 
when writing the transport equation
for the phonons one should keep in mind that the collective and thermal motion of the phonon fluid can
occur at long scales. For this reason one has to keep the Christoffel symbols in the transport equation,
which contain derivatives of the superfluid velocity. As it will be shown in Section~\ref{Sec-thermal},
the equilibrium distribution function of the phonons depends on ${\cal G}_{\mu \nu}$, meaning that
all the above quantities evaluated at equilibrium vary on scales of the order of the scale of variation of the acoustic metric.

From the expressions of the current, energy-momentum tensor
and entropy one can derive the continuity equations obeyed by these quantities. The analysis is quite simplified if one takes  into account the following property \cite{Lindquist}
\begin{equation}\label{magic}
\left [ \int p^{\mu_1} \cdots p^{\mu_l} p^{\nu} f(x,p) d {\cal P} \right]_{; \nu} =
\int p^{\mu_1} \cdots p^{\mu_l}  L[f] d {\cal P} \,.
\end{equation}
In the case of vanishing collision term, considering different values of $l$ in this equation  enables to obtain the various  covariant conservation laws. 

Taking $l=0$ one has that the covariant continuity equation for the phonon number is given by
\begin{equation}\label{covariantN}
 (n_{\rm ph}^\nu)_{;\nu} = \left [ \int  p^{\nu} f(x,p) d {\cal P} \right]_{; \nu} = \int  L[f] d {\cal P} = \int  C[f] d {\cal P} \, ,
\end{equation}
which one can write  explicitly as
\begin{equation}
 \partial_\nu n_{\rm ph}^\nu + \Gamma^\mu_{\mu \nu} n_{\rm ph}^\nu = \int  C[f] d {\cal P} \;,
\end{equation}
and expresses the fact that collisional processes  can violate phonon number. 
Moreover one has to consider interactions of the phonons with the superfluid background. This is reflected in the second term in the l.h.s of the above equation where a Christoffel symbol appears signaling that  propagation of phonons is taking place in  ``curved" space-time.
Since we can write
\begin{equation}
\Gamma^\mu_{\mu \alpha} = \frac{1}{\sqrt{- \cal G}} \partial_\alpha \sqrt{- \cal G}=
\frac{1}{c_s} \frac {\partial c_s}{\partial x^\alpha} \,,
\end{equation}
it follows that   in the collisionless case, the variation of the
number of phonons stems only from a possible space-time dependence
of the speed of sound.   Therefore in the collisionless case  the   covariant conservation law of the phonon current can be written  in a  compact form as
\begin{equation}
(n_{\rm ph}^\nu)_{;\nu} =\frac{1}{\sqrt{- \cal G}} \partial_\alpha \left( \sqrt{- \cal G} n^\alpha_{\rm ph} \right)  = \frac{1}{c_s} \partial_\alpha \left( c_s n^\alpha_{\rm ph} \right) =   0 \,.
\end{equation}
Notice that in the absence of collisions the quantity that is conserved with respect to the flat metric, i.e. with respect to the derivative $\partial_\alpha$, is the current
\be 
\tilde n^\alpha_{\rm ph} =\sqrt{- \cal G} n^\alpha_{\rm ph}\,.
\ee 
This is the quantity that should be identified  with the conserved phonon current, and obeys
the continuity equation
\be
\de_\alpha \tilde n^\alpha_{\rm ph} =0\,,
\ee
in the absence of collisions.

From Eq.~(\ref{magic}) with $l=1$, one obtains the covariant continuity equation for the energy-momentum
tensor
\be\label{covariantT}
(T^{\mu\nu}_{\rm ph})_{;\nu} = \int p^\mu  C[f] d {\cal P} = 0 \,,
\ee
where we have assumed that   collisions  conserve both energy and momentum.
The equation above can be written more explicitly as
\be
\label{covarianteqT}
(T^{\mu\nu}_{\rm ph})_{;\nu} =(T^{\mu\nu}_{\rm ph})_{,\nu} +
\Gamma^\nu_{\nu \alpha} T^{\mu \alpha}_{\rm ph} + \Gamma^\mu_{\nu \alpha} T^{\nu \alpha}_{\rm ph} =0 \,, 
\ee
that can be rewritten as follows,
\be\label{T1} \partial_\alpha \left( \sqrt{- \cal G} T^{\mu\alpha}_{\rm ph} \right) + \sqrt{- \cal G} T^{\nu \alpha}_{\rm ph} \Gamma^\mu_{\nu \alpha} = 0\,. \ee
As in the case of the phonon current,  it is useful to  redefine the energy-momentum tensor  as
\be\label{newTmunu}
 \tilde T^{\mu \nu} = \sqrt{- \cal G} T^{\mu \nu}\, ,
\ee
and in this way  the four-momentum associated with it is given  by
\be
P^{\nu}  =  \int d^3 x \, \sqrt{- \cal G}  T^{0 \nu} = \int d^3 x \,\tilde  T^{0 \nu} \,.
\ee
Notice that defining the  energy-momentum tensor  of the phonons as  $\tilde T^{\mu \nu}$, we have that the 
total conserved tensor of  the  system is given by 
\be
T^{\mu \nu}  =  T^{\mu \nu}_{0} +\tilde T^{\mu \nu}_{\rm ph} \,,
\ee
where $T^{\mu \nu}_0$ is the component of the energy-momentum associated to the superfluid background.

If we  assume that the collisional integral is zero, one can expect to have also the covariant conservation of entropy, i.e.
\begin{equation}
(s^\alpha_{\rm ph})_{;\alpha} = \frac{1}{\sqrt{- \cal G}} \partial_\alpha \left( \sqrt{- \cal G} s^\alpha_{\rm ph} \right)  = \frac{1}{\sqrt{- \cal G}}
\partial_\alpha \left( \tilde s^\alpha_{\rm ph} \right)=0 \,,
\end{equation}
where $\tilde s^\alpha_{\rm ph} =\sqrt{- \cal G} s^\alpha_{\rm ph} $. In the presence of collisions this relation is not satisfied,  and then there is entropy generation.

Let us remark on the meaning of the continuity equations of  phonons that we have derived 
\cite{Volovik:2000ua}. 
 Since the phonon fluid   moves on the top of the superfluid, even assuming that the collision term vanishes we still have the possibility that processes of interaction between the phonon fluid and the superfluid will lead to the exchange of momentum and energy among them. Indeed, even if energy and momentum of the whole system composed by the superfluid condensate and by the phonons is  conserved, there can be energy and momentum exchanges between these two subsystems. This is also what happens when considering 
matter propagating in a gravity field, as also in that case the energy-momentum tensor associated to the ``matter" field is not strictly conserved.

The possibility of exchanges of energy and momentum mentioned above 
is reflected in the continuity Eq.~(\ref{T1}) that we rewrite  in the compact form 
\be \label{conservationT2}
 \partial_\mu  \tilde T^\mu_{{\rm ph} \,\nu} \,  
= \,  \frac 12 \tilde T^{\mu \rho}_{\rm ph} \, \partial_\nu {\cal G}_{\rho \mu}  \,,
\ee
using the definition in Eq.(\ref{newTmunu}).
On  the right hand side of this  conservation law there is a term that couples the energy-momentum tensor of the phonons with  the derivative of the gravitational field, expressing in a covariant form the interaction between the phonons and  the underlying superfluid. As a result, energy and momentum can be exchanged between the phonons and the background  if the speed of sound is space-time dependent and/or if the superfluid velocity is space-time dependent. 
This fact can be understood also within the gravitational analogy.  Indeed, the energy-momentum tensor of any matter field, here the phonon field,    couples with the gravitational 
field ${\cal G}_{\mu\nu}$ inducing  exchange of energy and momentum  between the  gravitational and  matter  fields.

Let us finish this Section by stressing that the interesting  point of the transport theory 
presented here  is that the use of the  gravitational analogy 
automatically allows to write  the equations governing the variations of phonon currents and energy-momentum tensor in terms of covariant equations with respect to the acoustic metric.

\section{Thermal equilibrium of the superfluid phonons}
\label{Sec-thermal}


The covariant continuity equations for the phonons derived in the previous Section   take a very simple form when
the macroscopic variables associated to the  superfluid are homogeneous or constant. Indeed in this case
the Christoffel symbols vanish, i.e. $\Gamma^\mu_{\nu \alpha} =0$, and covariant derivatives are equal to  ordinary derivatives.
Thus, the kinetic equation for the phonons  are  the same that  one obtains for ordinary particles
evolving in a flat metric.
In this case one can find the equilibrium distribution function for the phonons in the
exact same way as for ordinary particles.

One can as well find global equilibrium solutions for stationary situations
(no time dependence). Global, or collisional equilibrium, distribution functions are given by
\begin{equation}
f_{\rm eq}(x,p) = \frac{1}{\exp{(p^\mu  \beta_\mu)} - 1} \, ,
\end{equation}
where $\beta_\mu$ is a vector that can be derived as follows. 
Since $L[f_{\rm eq}]= 0$ one has that
$\beta^\mu$ obeys the equation
\begin{equation}
\label{KVC}
\left( { \beta}_{\lambda, \rho} - { \beta}_\alpha \Gamma^\alpha_{\lambda \rho} \right) p^\lambda p^\rho =0 \,,
\end{equation}
that can be rewritten as
\begin{equation}
\label{equilib1}
{ \beta}_{\lambda; \rho} + { \beta}_{\rho;\lambda} = 0 \ .
\end{equation}
Assuming that $\beta_\mu$ is time independent, it is possible to check that one possible solution to
the above equations is given by
\begin{equation}
\beta^\mu = (1/T,{\bf 0}) \ ,
\end{equation}
where  $T$ is a constant that we identify with the temperature.

There are other local equilibrium configurations that are found by demanding no local entropy generation,
that is $s^\alpha_{; \alpha}=0$ 
 and with solutions given by the collisional invariants of the theory.
In this case we can expect that the equilibrium distribution function takes the form
\begin{equation}
\label{loc-eq}
f_{\rm eq}(x,p) = \frac{1}{\exp{(p^\mu  \beta_\mu)} - 1} \ ,
\end{equation}
where
\begin{equation}
\beta^\mu = \beta(x) u^\mu(x)  \,.
\end{equation}

In the local equilibrium configuration the phonon energy-momentum tensor has to take the form
\begin{equation}
T^{\mu \nu}_{\rm ph} =  H u^\mu u^\nu - K
{\cal G}^{\mu \nu} \, ,
\end{equation}
where $H$ and $K$ are two scalars. The reason for this form is  that $T^{\mu \nu}_{\rm ph}$ is a tensor and because of the definition in  Eq.~(\ref{var-T}) it can be built  by the tensors $u^\mu u^\nu$ and  ${\cal G}^{\mu \nu}$ only. We will fix the scalars $H$ and $K$ as follows.
When both the superfluid and
phonon fluid are at rest, {\it i.e.} $u^\mu = v^\mu = (1, {\bf 0})$, the pure temporal
component of this tensor should give the phonon energy density at rest, while the spatial components
are diagonal and proportional to the pressure
\begin{equation}
T^{00}_{\rm ph} = \rho_{\rm ph} \ , \qquad   T^{ij}_{\rm ph} = P_{\rm ph} \, \delta^{ij} \ .
\end{equation}
Thus one can identify
\begin{equation}
H = \rho_{\rm ph} +P_{\rm ph}/c_s^2 \ , \qquad K = P_{\rm ph} \,,
\end{equation}
and the phonon energy-momentum tensor is then expressed as
\begin{equation}\label{explicit1}
T^{\mu \nu}_{\rm ph} = (\rho_{\rm ph} +P_{\rm ph}/c_s^2) u^\mu u^\nu - P_{\rm ph} \eta^{\mu \nu}
- P_{\rm ph} \left(\frac {1}{c_s^2} -1 \right) v^\mu v^\nu \,.
\end{equation}

From the definition in  Eq.~(\ref{var-T}) one sees that  the energy-momentum tensor is  traceless, i.e. ${\cal G}_{\mu \nu} T^{\mu \nu}_{\rm ph} = 0$. Upon substituting the expression of $T^{\mu \nu}_{\rm ph}$ that  we have obtained  above, one has  that the traceless condition leads to the equation
\begin{equation}
\label{EoS-phonon}
\rho_{\rm ph} = \frac{P_{\rm ph}}{c_s^2} \left(\frac{4c_s^2 -{\cal G}_{\mu \nu} u^\mu u^\nu}{{\cal G}_{\mu \nu} u^\mu u^\nu}  \right) \,.
\end{equation}
Moreover, using the definition of the energy-momentum tensor in Eq.~(\ref{newTmunu}) 
and the condition ${\cal G}_{\mu \nu} T^{\mu \nu}_{\rm ph} = 0$, one finds 
\be\label{explicit2}
\tilde T^{\mu\nu}_{\rm ph} = 4 \tilde P_{\rm ph} \hat u^\mu \hat u^\nu  - \tilde P_{\rm ph}{\cal G}^{\mu\nu} \,,
\ee
where we have defined 
\be
\tilde P_{\rm ph}  =\sqrt{-{\cal G}} P_{\rm ph}  \qquad {\rm and } \qquad \hat u^\mu = \frac{u^\mu}{\sqrt{u^\mu  u^\nu {\cal G}_{\mu\nu}}}  \,.
\ee
We also define  $\hat u_\mu ={\cal G}_{\mu\nu}\hat u^\nu $ and therefore $\hat u^\mu \hat u_\mu = 1$, meaning that $\hat u^\mu $ is a  four-vector normalized to $1$ with respect to the metric ${\cal G}_{\mu\nu}$.

It is important to note that the phonon contribution to the energy-momentum tensor keeps the form
that is expected for the normal fluid component in a superfluid. Indeed, Eq.~(\ref{explicit1}) has
the tensorial structure which is expected in these systems according with the 
 Poisson bracket approach to the hydrodynamics  reviewed in the Appendix \ref{section-PB} (see Eq.~(\ref{SonTten})).

In a similar way one finds that the expressions of the current and of the entropy current are of the form
\begin{equation}
\tilde n^\mu_{\rm ph} =\tilde n_{\rm ph} u^\mu  \ , \qquad \tilde s^\mu_{\rm ph} =\tilde s_{\rm ph} u^\mu \,,
\end{equation}
where $\tilde n_{\rm ph}$ and $\tilde s_{\rm ph}$ are scalars.

Upon employing the definitions above in the  Eqs.~(\ref{var-n}), (\ref{var-T}) and
(\ref{var-S}), with the local equilibrium distribution function (\ref{loc-eq})
 one  can compute the   phonon current, entropy
current and pressure. Then the energy density can be deduced from  the pressure, according to Eq.~(\ref{EoS-phonon}).
The best strategy for evaluating the integrals in Eqs.~(\ref{var-n}), (\ref{var-T}) and
(\ref{var-S}) is  to choose the  frame where the superfluid is
at rest $v_\mu = (1,{\bf 0})$. Then, after defining $u^\mu = \gamma (1,
{\bf u})$, where $\gamma$ is the Lorentz factor, we obtain for $|{\bf u}|<c_s$
\begin{equation}
\tilde n_{\rm ph}  =  \frac{T^3 \zeta(3)}{\pi^2} \frac{1}{c_s^3 \gamma^4( 1 - \frac{{\bf u}^2}{c_s^2} )^2} \ , \qquad
\tilde  s_{\rm ph}  =  \frac{T^3 2 \pi^2}{45} \frac{1}{c_s^3 \gamma^4( 1 - \frac{{\bf u}^2}{c_s^2} )^2} \ , \qquad
\qquad
\tilde  P_{\rm ph}  =  \frac{T^4 \pi^2}{90 } \frac{1}{c_s^3 \gamma^4( 1 - \frac{{\bf u}^2}{c_s^2} )^2} \ .
\end{equation}

After realizing that in the superfluid rest frame
\begin{equation}
{\cal G}_{\mu \nu} u^\mu u^\nu = c_s^2 \gamma^2 ( 1 - \frac{{\bf u}^2}{c_s^2} ) \ ,
\end{equation}
one can  express the same quantities in any arbitrary frame writing
\begin{equation}\label{general}
\tilde n_{\rm ph}  = \sqrt{- {\cal G}} \frac{T^3 \zeta(3)}{\pi^2} \frac{1}{({\cal G}_{\mu \nu} u^\mu u^\nu)^{2}} \ , \qquad
\tilde s_{\rm ph}  = \sqrt{- {\cal G}} \frac{T^3 2 \pi^2}{45} \frac{1}{({\cal G}_{\mu \nu} u^\mu u^\nu)^{2}}\ , \qquad
\tilde P_{\rm ph}  = \sqrt{- {\cal G}} \frac{T^4  \pi^2}{90 } \frac{1}{({\cal G}_{\mu \nu} u^\mu u^\nu)^{2}} \,.
\end{equation} 
From these expressions one can recover  the non-relativistic results of  Ref.~\cite{IntroSupe}. Indeed, taking the limits
$u^\mu \rightarrow  (1,{\bf u}_{NR})$ and $v^\mu \rightarrow  (1,{\bf v}_{NR})$,
and using the same symbol for the non-relativistic speed of sound,  one obtains that 
\be
{\cal G}_{\mu \nu} u^\mu u^\nu \rightarrow c_s^2 - ({\bf u}_{NR}-{\bf v}_{NR})^2 \,,
\ee
and the expressions above turn into the
 relations
\begin{equation}
\tilde n_{\rm ph}  =  \frac{T^3 \zeta(3)}{\pi^2} \frac{1}{c_s^3 ( 1 - \frac{{\bf w}^2}{c_s^2} )^2} \ , \qquad
\tilde  s_{\rm ph}  =  \frac{T^3 2 \pi^2}{45} \frac{1}{c_s^3 ( 1 - \frac{{\bf w}^2}{c_s^2} )^2} \ , \qquad
\qquad
\tilde  P_{\rm ph}  =  \frac{T^4 \pi^2}{90 } \frac{1}{c_s^3 ( 1 - \frac{{\bf w}^2}{c_s^2} )^2} \,,
\end{equation}
where ${\bf w} = {\bf u}_{NR}-{\bf v}_{NR}$ is the so-called counterflow velocity \cite{IntroSupe}.

From the Eqs.(\ref{general}) one  can also verify that the thermodynamical  relation
\begin{equation}
\label{thermo-relation}
\tilde s^\mu_{\rm ph} =  \beta^\mu \tilde  P_{\rm ph}  + \beta_\nu \tilde T^{\mu \nu}_{\rm ph}  \,,
\end{equation}
 is satisfied. In this relation indices are lowered/raised with the acoustic metric.

A similar expression for the phonon contribution to the pressure in a cold relativistic superfluid
was found by Carter and Langlois \cite{Carter-Lang}. The main difference of our treatment with
that of Ref.~\cite{Carter-Lang} is that we assume that the velocity of the phonon fluid is
relativistic. Carter and Langlois assumed a non-relativistic velocity for the phonon fluid,
and accordingly, they obtained a non-relativistic relation among the thermodynamical variables
instead of the relativistic one reported in Eq.~(\ref{thermo-relation}).

For completeness we write the conservation law of the energy-momentum tensor  in Eq.(\ref{conservationT2})  in a more explicit form. Using the definition in Eq.(\ref{explicit2}) we can rewrite the left hand side of Eq.(\ref{conservationT2})  as 
\be
\partial_\mu  \tilde T^\mu_{{\rm ph}\,\nu}  = 4 [ (\de_{\mu} \tilde P_{\rm ph}) \hat u^\mu \hat u^\nu +  \tilde P_{\rm ph} (\de_{\mu}  \hat u^\mu) \hat u^\nu + \tilde P_{\rm ph} (\de_{\mu}  \hat u^\nu) \hat u^\mu ] - \de_{\nu} \tilde P_{\rm ph} \,,
\ee
while the right hand side of Eq.(\ref{conservationT2})  turns out to be
\be
\frac 12 \tilde T^{\mu \rho}_{\rm ph} \, \partial_\nu {\cal G}_{\rho \mu} = \frac 12 \tilde P_{\rm ph} (4 \hat u^\mu \hat u^\rho  - {\cal G}^{\mu\rho})\partial_\nu {\cal G}_{\rho \mu} \,.
\ee
Projecting these equations 	along $\hat u^\nu $ we obtain that
\be
 4  (\de_{\mu} \tilde P \hat u^\mu) -  \hat u^\mu \de_{\mu} \tilde P_{\rm ph} =  \frac{\tilde P_{\rm ph}}{2} (4  \hat u^\mu \hat u^\rho  - {\cal G}^{\mu\rho} ) u^\nu \partial_\nu {\cal G}_{\rho \mu} \,.
\ee
Projecting in the directions orthogonal to $\hat u^\nu $ with  $( \hat u^\alpha \hat u^\nu  - {\cal G}^{\alpha\nu}) $ one has 
\be
- 4  \tilde P_{\rm ph} {\cal G}^{\alpha \nu}  \hat u^\mu \de_{\mu} \hat u_\nu- (  \hat u^\alpha \hat u^\nu  - {\cal G}^{\alpha\nu}) \de_{\nu} \tilde P_{\rm ph} =  \frac{\tilde P_{\rm ph}}{2}  (4 \hat u^\mu \hat u^\rho  - {\cal G}^{\mu\rho})( \hat u^\alpha \hat u^\nu  - {\cal G}^{\alpha\nu})\de_\nu {\cal G}_{\rho \mu} \,.
\ee

In the limit where  the metric ${\cal G}_{\mu\nu}$ can be approximated with the flat metric $\eta_{\mu\nu}$, one has that  $\hat u^\mu \rightarrow u^\mu$ and the equations above reproduce the hydrodynamical equations of a ultrarelativistic system.

\section{Discussion}
\label{Discussion}


The properties of superfluids at low but non-vanishing temperature   are influenced  by the presence 
of  phonons, the massless Goldstone bosons originating from the spontaneous breaking of a global symmetry. 
In  case  these quasiparticles are the only low energy degrees of freedom they will
give the dominant  contribution to the transport properties of the system in this cold regime.

The transport theory of a non-relativistic superfluid  was developed long ago, see e.g. Ref.~\cite{IntroSupe}.  In the non-relativistic case  the  transport equations for the phonons can be expressed
as those of a boson evolving in Unruh acoustic metric \cite{Volovik:2000ua}. 
In the present paper we have extended this approach to relativistic superfluids.
We have shown that in a relativistic superfluid the same gravity analogy can be employed for deriving the transport equations replacing  Unruh's metric with a  generalized relativistic acoustic metric.  Similar results have been reported in   
Ref.~\cite{Popov:2006nc}, although implemented in a different manner.
The advantage of this formulation  is twofold. It allows us to  obtain the continuity
equations and to express the phonon hydrodynamical variables in a covariant form,  valid in an arbitrary reference frame. 
Moreover, it clarifies the physical meaning of the variables used in the convective variational
approach of the hydrodynamics.

The computation of transport coefficients can be  implemented employing  the framework presented in the present paper 
considering the Boltzmann equation (\ref{Blotzmman}) with a non-vanishing  collision term. The relevant collision terms    can be derived from the microscopic physics and depend on the particular system considered. For the CFL superfluid  in the high density limit the collision term can be evaluated considering various   scattering processes  among superfluid phonons whose vertices can be read  from the effective Lagrangian in Eq.~(\ref{L-BGB-0}). The leading processes are  binary collisions, collinear splitting and joining processes \cite{Manuel:2004iv}. The phonon contribution to the shear viscosity, and the bulk viscosity associated to the normal fluid component have been computed following this procedure in Refs.~\cite{Manuel:2004iv,Manuel:2007pz}. These computations were done  using kinetic
theory in the superfluid rest frame, where one takes $v^\mu =(1,{\bf 0})$, and
assuming homogeneity in the superfluid flow. This corresponds to consider in our equations vanishing Christoffel symbols, $\Gamma^\rho_{\mu \nu} =0$. The transport equation that we have presented will allow  us to compute the viscosity coefficients
that involve also  gradients in the superfluid flow. The results of these computations will be reported soon.

Although we have focused our attention on the CFL superfluid the  
techniques developed here  could be  employed  for different  relativistic superfluids.
In particular, it would be interesting to compute the phonon contribution to the different
viscosity coefficients in the neutron superfluid  that is supposed to be realized in the interior of
neutron stars. The evaluation of all the bulk viscosity coefficients arising from non-equilibrium
beta processes has been recently reported  in Ref.~\cite{Gusakov:2007px}. It would be interesting
to evaluate the phonon contribution to these viscosity coefficients.
One should also consider that at densities achievable in the core of neutron stars the energetically favored phase
might not be the CFL phase \cite{reviews}. Among the various possible  phases that can be realized one interesting possibility is that quark matter is in the Crystalline Color Suprconducting phase \cite{Bowers}, where the  gap parameter  is periodically modulated in space and therefore spontaneously breaks translational invariance. In this case the low energy degrees of freedom do not consist only of the phonon excitation associated to the spontaneous breaking of $U(1)_B$. Different  phonon excitations arising from the spontaneous  breaking of space symmetries and describing the oscillation of the crystalline structure are part of the  spectrum as well \cite{Casalbuoni}. Moreover,  gapless fermionic excitations  are present and this makes the analysis of the  low energy properties of this phase  challenging. In such a situation the transport theory derived in the present paper should be extended to include all the remaining low energy modes.

\begin{acknowledgments}
We thank Felipe Llanes-Estrada for useful discussions.
This work has been supported by the Ministerio de Educaci\'on y Ciencia (MEC) under grants
AYA 2005-08013-C03-02 and FPA2007-60275.
\end{acknowledgments}

\appendix

\section{Poisson bracket approach to the hydrodynamics of relativistic superfluids}\label{section-PB}

 The hydrodynamical equations of a relativistic superfluid have been derived 
using the Poisson bracket (PB) method by Son in Refs.~\cite{Son:2000ht,Son:unpubl}.  In this approach one considers that the  superfluid properties of a system 
arise from the  spontaneous breaking of a continuous $U(1)$ symmetry, with
 the appearance of a Goldstone mode. Since hydrodynamics
is an effective field theory valid at long time and long length scales,
 the standard fluid variables should couple to the Goldstone mode.

One defines the Hamiltonian of the system as a functional of the hydrodynamical variables and of the Goldstone field, and then postulates a set of PB  among them to  derive the hydrodynamical equations. 
The postulated brackets might be deduced from the canonical quantum commutators in the corresponding microscopic quantum theory. Some of them also follow from simple physical considerations.

The equation of state of the system can be deduced from the microscopic theory and  one finds that the
differential equation for the total pressure of the system  is given by
\begin{equation}
dP = s_0dT_0 + n_0 d\mu_0 + \frac 12 V^2 d(\partial_\mu \varphi)^2 \, ,
\end{equation}
where $V$ is a parameter depending on the microscopic variables. As an example  in complex scalar theories $V$ is 
 proportional to the  expectation value of  the scalar field \cite{Son:2000ht,Son:unpubl}.

After a Legendre transformation of the pressure, one  has that the energy density is given by $\rho= s_0T_0 + n_0 \mu_0 - P$.
Imposing Poisson brackets conditions on entropy density, particle density and momentum density one arrives at  the hydrodynamical equations \cite{Son:2000ht,Son:unpubl}
\begin{eqnarray}
\label{SonTten}
\partial_\rho T^{\rho \sigma} = 0 \ , \qquad T^{\rho \sigma} = \left(\rho + P\right)u^\rho u^\sigma -
P \eta^{\rho \sigma}
+ V^2 \partial^\rho \varphi \partial^\sigma \varphi \\
 \partial_\rho \left(n_0 u^\sigma - V^2 \partial^\rho \varphi\right) = 0 \\
\partial_\mu (s_0 u^\mu) = 0 \\
u^\mu \partial_\mu \varphi+ \mu_0 = 0 \,.
\end{eqnarray}
In this formulation of the superfluid hydrodynamics there is a clearer interpretation
of both the stress-energy tensor and current, as being due to the sum of the normal fluid
part and  the coherent motion of the condensate (the superfluid). Indeed both these quantities have contributions
proportional to  the hydrodynamical
velocity $u^\mu$, associated to the normal fluid, and to  the parameter $V$. Notice that the entropy has only contributions due 
to the normal component, as it should be.

While in principle this formulation of the hydrodynamics looks very different from the
one presented in Section~\ref{Sec-hydro} they are equivalent. Indeed it is possible to relate the various variables of the two approaches
by means of the following equations
\cite{Son:2000ht,Son:unpubl}
\begin{eqnarray}
\Lambda &=& \rho + V^2 (\partial \varphi)^2 \ , \qquad
\mu_\mu  =  - \partial_\mu \varphi  \ ,
\\  s^\mu & = & s_0 u^\mu \ , \qquad
n^\mu = n_0 u^\mu - V^2 \partial^\mu \varphi \ , \\
\Theta_\mu &=& \frac{1}{s_0} \left[ \left(T_0 s_0 + \mu_0 n_0\right) u_\mu + n_0 \partial_\mu \varphi \right] \,.
\end{eqnarray}

Even if the two approaches are very similar they differ in one subtle point. While in the PB approach
the value  of $u^\mu \partial_\mu \varphi$ is constrained, in the convective approach
this is considered as a free parameter. Instead one fixes the  norm of the gradient of the field $\varphi$, 
requiring that at $T=0$ one has $\partial_\mu \varphi \partial^\mu \varphi = \mu^2$.
Imposing the two normalizations simultaneously over-constraints the system,
reducing the degrees of freedom that the problem should have.
 Let us also comment that the quantity $\mu \neq \mu_0$. Indeed  $\mu_0$ should be considered as the
chemical potential associated to the normal component of the superfluid evaluated in the comoving frame.
Instead $\mu$ is the norm of the vector $\mu^\rho$.

\end{document}